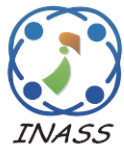

# MR-MOSLO: VM Consolidation Using Multiple Regression Multi-Objective Seven-Spot Ladybird Optimization for Host Overload Detection

**Akram Saeed Aqlan Alhammadi[1]***     **Vasanthi Varadharajan[2]**

[1]*Computer Science Department, Rathinam College of Arts and Science,
Bharathiar University, Coimbatore, India*
[2]*Department of Information Technology, Sri Krishna Adithya college of Arts and Science,
Bharathiar University, Coimbatore, India*
* Corresponding author's Email: Akram_aqlan@hotmail.com

**Abstract:** Virtual Machine (VM) consolidation is a crucial process in improving the utilization of the resource in cloud computing services. As the cloud data centers consume high electrical power, the operational costs and carbon dioxide releases increases. The inefficient usage of the resources is the main reason for these problems and VM consolidation is a viable solution. VM consolidation includes host overload/under-load detection, VM selection and VM placement processes. Most existing host overload/under-load detection approaches of VM consolidation uses CPU utilization only for the determining host load. In this paper, three resources namely CPU utilization, memory utilization and bandwidth utilization are used for host overload detection and an adaptive regression based model called Multiple Regression Multi-Objective Seven-Spot Ladybird Optimization (MR-MOSLO) is proposed. This model is based on combining the benefits of adaptive threshold based and regression based host overload detection algorithms. This approach of combining these features provide more advantages for threshold setting in dynamic environments with accurate prediction of host overloading. For this purpose, initially, Multiple Regression (MR) algorithm is used which relay on CPU utilization, memory utilization and bandwidth utilization for estimation of the host load conditions. Then a Multi-Objective Seven-Spot Ladybird Optimization (MOSLO) algorithm is introduced to select the upper and lower threshold limits for host utilization. Based on these algorithms, the host overload/under-load is detected with high accuracy and less power consumption. The simulations are conducted in CloudSim tool and the empirical results shows that the proposed MR-MOSLO algorithm detects the host overload efficiently. The results obtained for 25 hosts, 30 VMs and 500 tasks, are: SLATAH is 20.0434, PDM is 8.7E-4, SLAV is 3.7E-5 and ESV is 10.962 which are lesser than the other methods. Though the energy of 15.4 kWh and SLA of 0.00757 are negligibly higher than some of the existing methods, the proposed approach provided comparatively better performance.

## 1. Introduction

Cloud computing has developed into the most reliable and popular computing model. It provides interconnected computers to provision the computing resources based on the SLA (Service Level Agreement) between cloud users and service providers [1]. The cloud computing data centers consume high amount of energy and leads to high power costs and increased carbon dioxide emissions [2]. The inefficiency in utilizing the resources is one of the major reasons for these problems [3]. To overcome such utilization problems, the dynamic provisioning of resources using VM consolidation is a best option. The VM technologies provide opportunity for consolidation and environment isolation. VM consolidation consists of host overload/under-load detection, VM selection and VM placement processes for live VM migration [4].

VM consolidation approaches perform live migration of VM to optimize the resource utilization of the cloud data centers [5, 6]. It migrates the VMs into smaller number of active physical machines (PMs) in order to turn the VM-less PMs into idle state to reduce the energy wastage. However, performing VM consolidation aggressively can lead to performance degradation due to increased response time or resource failures when an application may encounter unexpected resources requirements. Also, performing the VM migration for VM consolidation may cause SLA violations. This will further reduce the reliable Quality of Service (QoS) for cloud providers [7]. Therefore the algorithms for VM consolidation must be developed that not only cut down power consumption but also attends preferred QoS and promising the SLA. Many researches have focused on developing such features based VM consolidation but most host overload detection approaches are either static threshold [8] or adaptive threshold based methods [9, 10]. The static methods are not suitable for dynamic provisioning while the adaptive methods are suitable for dynamic provisioning but lacks accurate prediction. Likewise the other major issue is that most approaches rely on only one parameter, i.e. CPU utilization in determining the VM load. This approach also reduces the overall effectiveness of the host overload detection algorithm [11].

This paper considers the issues of the existing models and has aimed to develop an efficient VM consolidation approach with effective host overload detection. In this approach, the multiple regression and multi-objective seven-spot ladybird optimization algorithm [12] is proposed to detect the host overload and under-load conditions. The proposed approach calculates the resource utilization parameters namely CPU utilization, memory utilization and bandwidth utilization for the hosts. Then the multiple regression concepts are applied to determine the host utilization from the VM utilization. Finally, the multi-objective optimization using MOSLO is introduced to detect the overload or under-load condition of the host. Unlike existing methods, this approach is adapted to the dynamic nature of resource utilization. This approach also avoids the SLA violations by altering the overloaded hosts in an instant manner such that the required utilization is lesser than the actual utilization capacity. In addition to the overload detection in the current hosts, the new destination host list will also be certified to place the VMs from overloaded hosts. The experimental results convey that the proposed approach improves the adaptive threshold concept and also presents a highly energy efficient VM consolidation.

The rest of the article is organized as follows: Section 2 presents a brief discussion of the recent related works. Section 3 presents the formulation of the dynamic VM consolidation and section 4 demonstrates the proposed multiple regression and MOSLO algorithm. The experimental results are highlighted in section 5 while the conclusion of this paper is provided in section 6.

## 2. Related works

VM consolidation is a trending research topic and many researches have been focused on improving the performance of VM consolidation process. Arianyan et al. [13] introduced a VM consolidation method based on energy and SLA efficient resource management heuristics based on multi-criteria decision making method. This approach performs both the detection of under-loaded hosts and migrated VMs placement. This approach reduces the energy consumption, SLA violation, and number of VM migrations. The limitation of this model is that it considers only the CPU utilization for VM consolidation. Sharma and Saini [14] introduced a VM consolidation using median based threshold approach. This approach meets the SLA and also deals with energy-performance trade-off for the auto-adjustment of lower and upper threshold values for dynamic consolidation of VMs. This approach provided minimum level SLA with minimum performance degradation and same energy consumption. The limitation is that the power consumption is still slightly higher.

Mohiuddin and Almogren [15] presented a workload aware VM consolidation method using optimization of load of the server and cost of migration. This approach minimized the energy consumption, less heat generation and resource wastage but also improved the processing speed. However, this model has relatively low performance when the energy source has less power supply. Sharma et al. [16] proposed a failure-aware energy-efficient VM consolidation approach based on exponential smoothing. This approach considers the failure occurrence and threat rate of physical resources for VM consolidation along with VM resource management policies. This approach reduces the energy wastage and increased the reliability with high fault tolerance and less computation time. However, the presence of correlation failures is not considered in the approach. Lianpeng Li [17] presented SLA-aware and energy-efficient VM consolidation using robust linear regression prediction model. This model utilized error to amend the prediction errors and reduced the power consumption and SLA violation. This model employed the adaptive lower utilization threshold based on the interquartile range to determine the host under loading. Empirical results showed that this model reduced energy consumption by 25.43% and SLA violations by 99.16%. However, this VM consolidation model considers only the CPU utilization and does not consider the other resources RAM and network bandwidth.

Li et al. [18] proposed an energy-efficient and quality-aware VM consolidation using a discrete differential evolution algorithm. This approach optimizes the energy efficiency and service quality by detecting a global optimum solution for VM placement problem. This model reduces energy consumption, avoids unnecessary host overloading risk, and improves QoS. However, this model considers only the energy efficiency and service quality for VM placement but the major feature of consolidation i.e. maximizing resource utilization is not given high priority. Li et al. [19] proposed an energy-aware dynamic VM consolidation (EC-VMC) method for VM migration based on checks on the possibilities of multiple types of resources being overloaded. This approach utilized multiple algorithms for different phases of VM dynamic consolidation and used artificial bee colony foraging behavior to find the mapping relation between PMs and VMs for feasible solution. The results provided better performance of VM consolidation with efficient performance metrics and maximized utilization and guaranteed QoS. Li et al. [20] also presented a Bayesian network-based VM consolidation for live VM migration based on energy consumption and QoS parameters like the dynamic workload, CPU utilization and number of VM migrations. This approach minimizes energy consumption, avoids extra insignificant VM migrations, and improves QoS along with reducing inefficient resource consumption. However, both these methods consider the CPU utilization only and avoid other resources.

Ranjbari and Torkestani [21] introduced learning automata-based algorithm for VM consolidation with energy efficiency and SLA guarantee. This algorithm considers changes in the user demanded resources to predict the overloaded PM and shuts down idle servers to reduce the energy consumption. However, this model only detects the host overload while the underutilized hosts are not effectively detected. Mahdhi and Mezni [22] proposed a prediction-based VM consolidation approach in which the Kernel Density Estimation technique is used to predict the future VM migration traffic and resources. Using this approach, the energy consumption is minimized and QoS are guaranteed. This is one of the approaches that consider CPU, RAM and Storage for VM consolidation. However, the traffic and security risks are high in this model.

In a similar approach, Abdelsamea et al. [23] presented a VM consolidation approach using hybrid regression algorithms named Multiple Regression Host Overload Detection (MRHOD) and Hybrid Local Regression Host Overload Detection algorithm (HLRHOD), which considers the CPU, RAM and network bandwidth. This approach provided two models to detect the host overload and under-load effectively with reduced energy consumption and guaranteed QoS. However, it uses a regression formula to normalize the predicted utilization with a fixed triggering point which might degrade the effective performance.

In [24], the authors have developed five adaptive heuristics for host overload detection. i) Inter Quartile Range (IQR) decide the threshold of a host to be marked as overloaded using interquartile range, ii) Median Absolute Deviation (MAD) estimates median absolute deviation value to detect the overload condition, iii) Threshold method (THR) provides threshold for a host detect overloaded host; iv) Local Regression (LR) and v) Local Robust Regression (LRR) utilizes regression models to predict the host utilization for overload detection. These five heuristic methods also rely on mean and standard deviation of resource utilization to estimate the future load of a VM for overload detection. However, mean and standard deviation are influenced by terminal values which are static and less efficient for systems with dynamic resource utilization.

Similarly, another method called Mean, Median and Standard deviation [MMSD] based Overload Detection is presented in [25]. This method employs the median in addition to mean and standard deviation to determine host overload. However, this algorithm also suffers from the problem of terminal values.

From the literature, it can be inferred that most VM consolidation techniques utilize only the CPU utilization for detecting the overload and under-load conditions. This approach is easier to compute but there are drawbacks in effective prediction of load conditions when relying on single objective. Another important problem is the use of ineffective threshold methods for determining the overload thresholds. These problems are considered as the motivations for developing the proposed MOSLO based multiple regression algorithms for host overload and under-load detection approach.

## 3. Dynamic VM consolidation and VM selection

Dynamic VM consolidation is the effective strategy to reduce the energy depletion by dynamically varying the number of active VMs and PMs based on the user resource demands. This problem is formulated in MR-MOSLO as a multiple regression based optimization problem. The host load conditions based on CPU, RAM and bandwidth resources are estimated using the multiple regression model. The predicted host utilization value determines the host overload condition but determining a fixed threshold will not be suitable for the dynamic computing environment. The dynamic host overload detection is formulated to support both the optimization and regression solutions. It can be modelled as

$$Y = \frac{w_1}{1-CPU} \times \frac{w_2}{1-RAM} \times \frac{w_3}{1-BW} \qquad (1)$$

Where $w_i$ denotes the weight of CPU, RAM and bandwidth whose values are in the range of [0,1], $CPU$ denotes the physical host CPU utilization, $RAM$ denotes the physical host memory utilization, and $BW$ denotes the physical host bandwidth utilization. By computing these values, the regression coefficients can be estimated and used to predict the future host utilization. The predicted utilization value should be accommodated by upper and lower threshold values that are determined optimally using the MOSLO algorithm.

Based on these predicted values, the host overload is detected. If the host is overloaded, the next step will be the migration of the VMs to reduce the performance degradation. However, the selection of VMs requires specific approaches. Maximum correlation is the most utilized selection model and hence it is used in this paper. Maximum correlation approach estimates the correlation between CPU utilizations and selects the VM with higher correlation for live migration. Once selecting a VM to migrate, the host is checked for overload at each iteration. If it is still overloaded, the selection approach again selects the next VM for migration. This process will be repeated until the host reaches a state where it is not overloaded.

## 4. Multi-objective seven-spot ladybird optimization based multiple regression for host overload detection

The proposed MR-MOSLO algorithm performs the host overload detection in two stages. In the first stage, the multiple regressions are applied to estimate the predicted utilization of the host. In the second stage, the MOSLO algorithm optimally determines the upper and lower thresholds for predicted utilization. The general multiple regression model contains more than one regression variable or coefficients. This is modelled as the Ordinary Least Square (OLS) equation that contains the parameters of interest.

$$y = b_0 + b_1 x_1 + b_2 x_2 + \cdots + b_k x_k \qquad (2)$$

Minimization of this equation in terms of ordinary least squares criterion can provide the regression coefficients. It can be minimized with respect to the k-variables of the given system [23].

$$\min S = \sum_{i=1}^{n}(y_i - b_0 - b_1 x_{i1} - b_2 x_{i2} - \cdots - b_k x_{ik})^2 \qquad (3)$$

Modelling the problem of dynamic VM consolidation host overload detection, the x variables are replaced by the CPU, RAM and BW. Primarily, the CPU, RAM and BW utilization values of each host is estimated as the division of the average utilizations of all VMs in the host by the maximum utilization of host. Then the multiple regression algorithms will have matrices of X and $C$ where $C$ matrix is formed based on the values obtained from Eq. (1) for each host. For a system where i = 1,2,3; the multiple regression problem can be modelled in matrix form based on multiple regression with three independent variables [25].

Input multi-dimensional matrix
$$X = \begin{bmatrix} CPU & RAM & BW \\ x_{11} & x_{12} & x_{13} \\ x_{21} & x_{22} & x_{23} \\ x_{31} & x_{32} & x_{33} \end{bmatrix}$$

$$\text{Output matrix } C = \begin{bmatrix} \sum \chi_1\, y \\ \sum \chi_2\, y \\ \sum \chi_3\, y \end{bmatrix}$$

$$\text{Coefficient matrix } B = \begin{bmatrix} b_1 \\ b_2 \\ b_3 \end{bmatrix} = C \times X^{-1} \qquad (4)$$

From these matrices, the regression coefficients $B = (b_1, b_2, b_3)$ can be computed. The values of $b_0$ are obtained by the following equations:

$$b_0 = \bar{y} - b_1\overline{x_{i1}} - b_2\overline{x_{i2}} - b_3\overline{x_{i3}} \quad (5)$$

Where $\bar{y}$ is the mean of Y variables, $\overline{x_{i1}}$ is the mean of CPU utilization, $\overline{x_{i2}}$ is the mean of RAM utilization and $\overline{x_{i3}}$ is the mean of Bandwidth utilization. Based on the calculated regression coefficients, the predicted host utilization equation can be formed as

$$\begin{aligned}predictedutilization = \\ b_0 + (b_1 \times CPUutilization) + (b_2 \times \\ RAMutilization) + (b_3 \times BWutilization)\end{aligned} \quad (6)$$

By replacing values of CPU utilization, RAM utilization and BW utilization in the equation, the predicted Utilization is obtained. Then the thresholds are determined for this predicted Utilization based on the MOSLO algorithm. The optimal lower threshold $Th_{low}$ and upper threshold $Th_{upper}$ limits are determined based on current host utilizations (CPU, RAM and BW).

The MOSLO algorithm optimally selects the threshold levels by iteratively running the cloud host and estimating the utilization in 'n' runs. These utilization values are ordered by the algorithm by comparing the values of CPU, RAM and BW with CPU establishing the higher priority. Hence based on the CPU utilization as the primary sorting criteria, the host utilization values are ranked in descending order. From this ranking order, the highest ranking utilization value except 1 or 100% is taken as $Th_{upper}$ while the smallest utilization value apart from 0 is taken as $Th_{low}$. These thresholds are used in the host overload detection by comparing with the predicted utilization from MR. these threshold values are computation for each fresh new operation of the cloud system.

MOSLO is based on the foraging behavior of seven-spot ladybirds and can solve the above mentioned threshold selection problem. Initially, the search space of the utilizations obtained is determined. Then the search space is divided into subspaces (patches) $n$ for easier comparison. Then the initialization population of seven-spot ladybirds are treated as a utilization value in the subspaces. Considering the number of seven-spot ladybirds as $m$ positioned randomly in the subspace, then the complete population size will be denoted by $N$, $N = m \times n$. Then the fitness is calculated for each ladybird using the following equation:

$$\min F(X) = |W_1 \times f_1(x), W \times f_2(x), W \times f_3(x)| \quad (7)$$

Where $f_1(x) = CPU$; $f_2(x) = RAM$; $f_3(x) = BW$; $W_1$ is the high weight parameter and W is the smaller weight parameter. The CPU utilization is given high weight parameter to initiate its high priority in the ranking order determination. The fitness function is first decomposed into individual single-objective sub-problems which are solved in a collaborative manner to form the objective values.

The estimated current fitness values are compared with the other fitness to evaluate the best. First, the fitness of each ladybird is matched with best historical position denoted as *s*best. When current fitness is better than *s*best, the original *s*best value and its position are replaced by the current fitness. When comparing the current fitness of all ladybirds in a patch with their preceding best position denoted as *l*best, the better fitness takes the place of *l*best or else it stays the same. Likewise, the comparison between fitness of all ladybirds in a population with their preceding best position denoted as *g*best results in the better fitness positioned as *g*best between the two values.

If the position of ladybird does not improve after T iterations, then new positions are produced in each patch and abandon the old position. The new position is produced near the *g*best to share the information of the best ladybird as follows:

$$p'_{i,j} = p_{gbest,j} + \phi\omega \quad (8)$$

Where $p'_{i,j}$ is the new position, $p_{gbest,j}$ is the old position, $\omega$ is the neighbourhood space of *g*best and $\phi$ is a random number between [−1, 1].

The position of the updated at each iteration based on the velocity. This phenomenon is performed in two searches: the extensive search which is slow movement and the intensive search is the faster and linear movement. The update equation after the extensive search is given by

$$V_i(t) = c \times r_1 \times (S_i(t) - P_i(t)) + \varepsilon_1 \quad (9)$$

$$P_i(t+1) = P_i(t) + V_i(t) \quad |V_i(t)| \leq V_{max} \quad (10)$$

The update equation after the intensive search is given by

$$V_i(t) = c \times r_2 \times (L_i(t) - P_i(t)) + \varepsilon_2 \quad (11)$$

$$P_i(t+1) = P_i(t) + V_i(t) \quad |V_i(t)| \leq V_{max} \quad (12)$$

Where $r_1$ and $r_2$ are the random values between [0, 1]; c is the positive constant to adjusting the step size and search direction in each iteration. $V_i(t)$ is the velocity, $P_i(t)$ is the current position, $P_i(t+1)$ is the newly moved position, $V_{max}$ is the maximum velocity computed based on upper and lower bounds of each patch, $S_i(t)$ is the new position moved away from gbest, $L_i(t)$ is the new position moved towards gbest, $\varepsilon_1$ and $\varepsilon_2$ are relatively smaller random values set to adjust the position of ladybirds i.e. to round off the decimal values of the calculated fitness values.

Once this step is completed, the two threshold limits $Th_{low}$ and $Th_{upper}$ are returned. Or else, the process is again repeated by reshuffling and recalculating the fitness until the solutions are obtained. Thus the thresholds are obtained and used to determine the host overload condition. When the predicted utilization is between $Th_{low}$ and $Th_{upper}$, the corresponding host is considered as normal loaded. Meanwhile, if the predicted utilization is less than $Th_{low}$, the host is under-loaded and if the predicted utilization is greater than $Th_{upper}$, the host is considered as overloaded. Algorithm 1 summarizes the complete processes of MR-MOSLO.

**Algorithm 1: MR-MOSLO**
Input: CPU utilization, RAM utilization, BW utilization
Output: Decision on whether host is overloaded, under-loaded or normal
For each host in host list do
    For each VM do
        $X \leftarrow$ Multidimensional matrix {CPU, RAM, BW}
        $Y \leftarrow \frac{w_1}{1-CPU} \times \frac{w_2}{1-RAM} \times \frac{w_3}{1-BW}$
    Apply OLS Multiple Linear regression using Eq. (2), (3);
    Compute the regression coefficients using Eq. (4)-(5);
    Estimate the predicted utilization using Eq. (6)
    Apply MOSLO
    Initialize SLO and design search space;
    Iteration T=0;
    Set ladybirds as utilization values;
    For each ladybird
        Compute host utilization in each iteration;
        Estimate fitness values using Eq. (7)
        Compare & determine sbest, lbest and gbest;
        Rank utilization in descending order using CPU as priority
        If no position = sbest, lbest or gbest
        Produce new position using Eq. (8)
        Else if
        Update positions
        T=T+1;
            If extensive search
            Update positions using Eq. (9), (10);
            Else if intensive search
            Update positions using Eq. (11), (12);
            End if
        Return $Th_{low} \neq 0$ and $Th_{upper} \neq 1$;
End for
If $predicted utilization \geq Th_{upper}$,
    Host is overloaded;
Else If $predicted utilization \leq Th_{low}$,
    Host is under-loaded;
Else If $Th_{low} > predicted utilization > Th_{upper}$,
    Host is Normal loaded;
    Repeat process from beginning;
    End if
    End for
End for

## 5. Experimental results

The performance of the proposed MR-MOSLO host overload detection algorithm for VM consolidation is evaluated in CloudSim tool. We have considered 800 heterogeneous physical nodes as servers. These servers are assigned with 1860MIPS (Million instructions per second) for the core. Network bandwidth is considered as 1GB/s and VMs are single core. Data centers include 100 hosts and 100 VMs using random workload traces. Hosts of two types namely HP ProLiant ML110 G4 (Intel Xeon 3040, 2 cores 1860 MHz, 4 GB), and HP ProLiant ML110G5 (Intel Xeon 3075, (2 cores 2660 MHz, 4 GB) are used. The host overload is frequently evaluated according to the scheduling interval set to 300 seconds.

The performance of this algorithm is compared with the existing host overload detection algorithms presented in [23-25] namely THR, IQR, MAD, LR and LRR [24]; MMSD [25] and MRHOD and HLRHOD [23]. The parameters used for comparison are Total energy consumption, SLA, SLATAH, PDM, SLAV and ESV parameters which are explained in [26, 27].

Table 1. Performance evaluation (25 hosts & 500 tasks)

| Parameter | 30 | 50 | 70 | 90 | 110 | 130 | 150 | 170 | 190 |
|---|---|---|---|---|---|---|---|---|---|
| Energy (kWh) | 15.4 | 28.0 | 34.0 | 48.0 | 46.77 | 59.15 | 71.98 | 77.03 | 88.812 |
| SLA | 0.00757 | 1.167 | 1.167 | 15.43 | 48.59 | 42.98 | 37.29 | 32.767 | 33.365 |
| PDM | $8.7 \times 10^{-4}$ | 0.233 | 0.166 | 2.204 | 6.718 | 5.03 | 4.324 | 4.8153 | 4.3861 |
| SLATAH | 20.0434 | 20.175 | 20.443 | 20.125 | 38.671 | 34.37 | 31.90 | 27.65 | 17.282 |
| SLAV | 3.7E-5 | 0.567 | 0.567 | 0.275 | 189.59 | 167.9 | 146.09 | 133.16 | 75.804 |
| ESV | 10.962 | 18.58 | 28.58 | 40.47 | 37.156 | 33.423 | 310.17 | 275.70 | 673.23 |

## 5.1 Performance metrics

i. Total energy consumption (E) is the amount of electricity used by all resources in a data center and it is measured in kilowatt-hours (kWh).

ii. Service Level Agreement (SLA) is an agreement between a cloud service provider and its customers which can be determined by characteristics as minimum throughput or maximum response time. It can be estimated as 100 divided by the percentage of the application performance done at any time.

iii. SLA violation Time per Active Host (SLATAH) is the percentage of time when the host stays overloaded (CPU utilization is 100%).

$$\text{SLATAH} = \frac{1}{N}\sum_{i=1}^{N}\frac{T_{oi}}{T_{ai}} \quad (13)$$

where $N$ is the number of hosts, $T_{oi}$ is the total time when the host $i$ has 100% utilization, $T_{ai}$ is the total time when the host $i$ is in active state.

iv. Performance Degradation due to Migrations (PDM) is the estimated performance degradation due to the VM migrations.

$$\text{PDM} = \frac{1}{M}\sum_{i=1}^{M}\frac{C_{di}}{C_{ri}} \quad (14)$$

where $M$ is the number of VMs, $C_{di}$ is the estimated performance degradation of the VM $i$ due to migrations, $C_{ri}$ is the total CPU capacity.

v. SLA violation (SLAV) is the measure of violations occurred that effects the service experience of the clients. It is estimated as the product of SLATAH and PDM.

$$SLAV = SLATAH \times PDM \quad (15)$$

vi. Energy and SLA Violations (ESV) is measured as the product of Total energy consumption (E) and Service level agreement violation (SLAV).

$$ESV = E \times SLAV \quad (16)$$

## 5.2 Results and discussion

Table 1 shows the performance metrics obtained for the proposed MR-MOSLO when the number of hosts is set as 25 and number of Tasks is set as 500 while the number of VMs is varied from 30 to 190. Table 2 shows the performance metrics obtained for the proposed MR-MOSLO when the number of hosts is set as 800 and number of Tasks is set as 2500 while the number of VMs is varied from 30 to 190. Table 3 shows the comparison of the proposed MR-MOSLO with the existing host overload detection algorithms.

Reduced energy consumption is the main objective for developing the proposed VM consolidation approach. When the number of hosts is set as 25 and number of Tasks is set as 500, the energy consumption for 20 VMs is 15.4 kWh which is significantly less. The increase in number of VMs generally increases the overall energy consumption. Even though the VMs are increased, the energy consumption is less compared to the ratio of VMs. It is also seen from Table 2 that when the number of hosts is set as 800 and number of Tasks is set as 2500, the results of MR-MOSLO are quite promising. From Table 3 it can be seen that the energy consumption of MR-MOSLO is lesser than six of the compared algorithms. Only MRHOD and HLRHOD have lesser energy consumption than the proposed approach. Although two methods have better energy consumption, considering the fixed triggering point in MRHOD and HLRHOD compared to the adaptive triggering point in MR-MOSLO, it can be negligible.

Table 2. Performance evaluation (800 hosts & 2500 tasks)

| Parameter | 30 | 50 | 70 | 90 | 110 | 130 | 150 | 170 | 190 |
|---|---|---|---|---|---|---|---|---|---|
| Energy (kWh) | 104 | 104 | 104 | 104 | 104 | 104 | 104 | 104 | 104 |
| SLA | 22.963 | 22.963 | 22.963 | 22.963 | 22.963 | 22.963 | 22.963 | 22.963 | 22.963 |
| PDM | 22.066 | 19.614 | 17.653 | 16.048 | 15.35085 | 14.71123 | 14.71123 | 13.57960 | 11.76898 |
| SLATAH | 0.32035 | 0.3203 | 0.3203 | 0.3203 | 0.32035 | 0.32035 | 0.32035 | 0.3203 | 0.3203 |
| SLAV | 7.06922 | 6.2837 | 5.6553 | 5.1412 | 4.917719 | 4.712814 | 4.712814 | 4.350290 | 3.770251 |
| ESV | 73519.9 | 65351 | 58815 | 53469 | 51144.27 | 49013.26 | 49013.26 | 45243.01 | 39210.61 |

Table 3. Performance comparison MR-MOSLO vs. other algorithms

| Algorithms | Energy | SLA | PDM | SLATAH | SLAV | ESV |
|---|---|---|---|---|---|---|
| THR [24] | 41.81 | 0.03048 | 0.23 | 12.99 | 2.987 | 124.917 |
| IQR [24] | 36.4 | 0.06521 | 0.27 | 20.85 | 5.629 | 204.914 |
| MAD [24] | 37.84 | 0.04304 | 0.25 | 17.34 | 4.335 | 164.036 |
| LRR [24] | 19.7 | 0.00765 | 0.031 | 99.12 | 3.001 | 59.12 |
| LR [24] | 19.7 | 0.00765 | 0.031 | 99.12 | 3.001 | 59.12 |
| MMSD [25] | 34.57 | 0.01921 | 0.09 | 20.45 | 1.841 | 63.626 |
| HLRHOD [23] | 13.53 | 0.00744 | 0.01 | 82.05 | 0.82 | 11.101 |
| MRHOD [23] | 13.48 | 0.0066 | 0.01 | 67.67 | 0.804 | 10.8406 |
| MR-MOSLO | 15.4 | 0.00757 | $8.7 \times 10^{-4}$ | 20.0434 | $3.7 \times 10^{-5}$ | 10.0962 |

SLAV is another key parameter that indicates the QoS of the system. SLAV is depended on PDM and SLATAH and when considering SLA of the system, SLAV should be lesser if not completely avoided. From Tables 1 and 2, it can be seen that the SLAV value is fluctuating for different number of VMs. When the number of hosts is 25 and number of Tasks is 500, the SLAV obtained is 3.7E-5, 0.567, 0.567, 0.275, 189.59, 167.9, 146.09, 133.16, and 75.804. The inconsistency in the SLAV is due to the frequent changes in the number of active hosts due to VM migration. Considering both scenarios, the minimum SLAV obtained is 3.7E-5 and maximum SLAV is 189.59 which can be termed significantly better considering the existing algorithms. Comparing the results with existing algorithms in Table 3, the proposed MR-MOSLO has considerable lesser SLAV as the proposed overload detection method has predicted the overloaded host efficiently and avoided unnecessary VM migrations.

The PDM and SLATAH parameters are also minimal as the proposed approach reduced SLA violation due to overload and associated energy to a considerable level. These advantages are justified by the performance results.

The ESV is the metric that provides the relationship between the Energy consumption and SLA violation. Both energy and SLAV metrics are significantly impact one another. From Tables 1 and 2, it is inferred that the ESV metric of the proposed MR-MOSLO has significantly improved over the increasing number of VMs. When comparing with the existing algorithms in Table 3, ESV of the proposed approach is 10.0962 which is lesser than all the other compared algorithms. Although the energy consumption of MR-MOSLO is higher than MRHOD and HLRHOD algorithms, the reduced SLAV has influenced the ESV metric.

The comparison results are plotted in Fig. 1 and it is inferred that the proposed MR-MOSLO has better performance in most parameters. Although not entirely superior, the proposed MR-MOSLO provides comparatively better performance. This can be a significant improvement in VM consolidation process. Thus the proposed MR-MOSLO provides comparatively better performance in host overload detection with high accuracy, almost equal energy and SLA values while comparatively lesser SLATAH, PDM, SLA violations and ESV values and improved QoS performance.

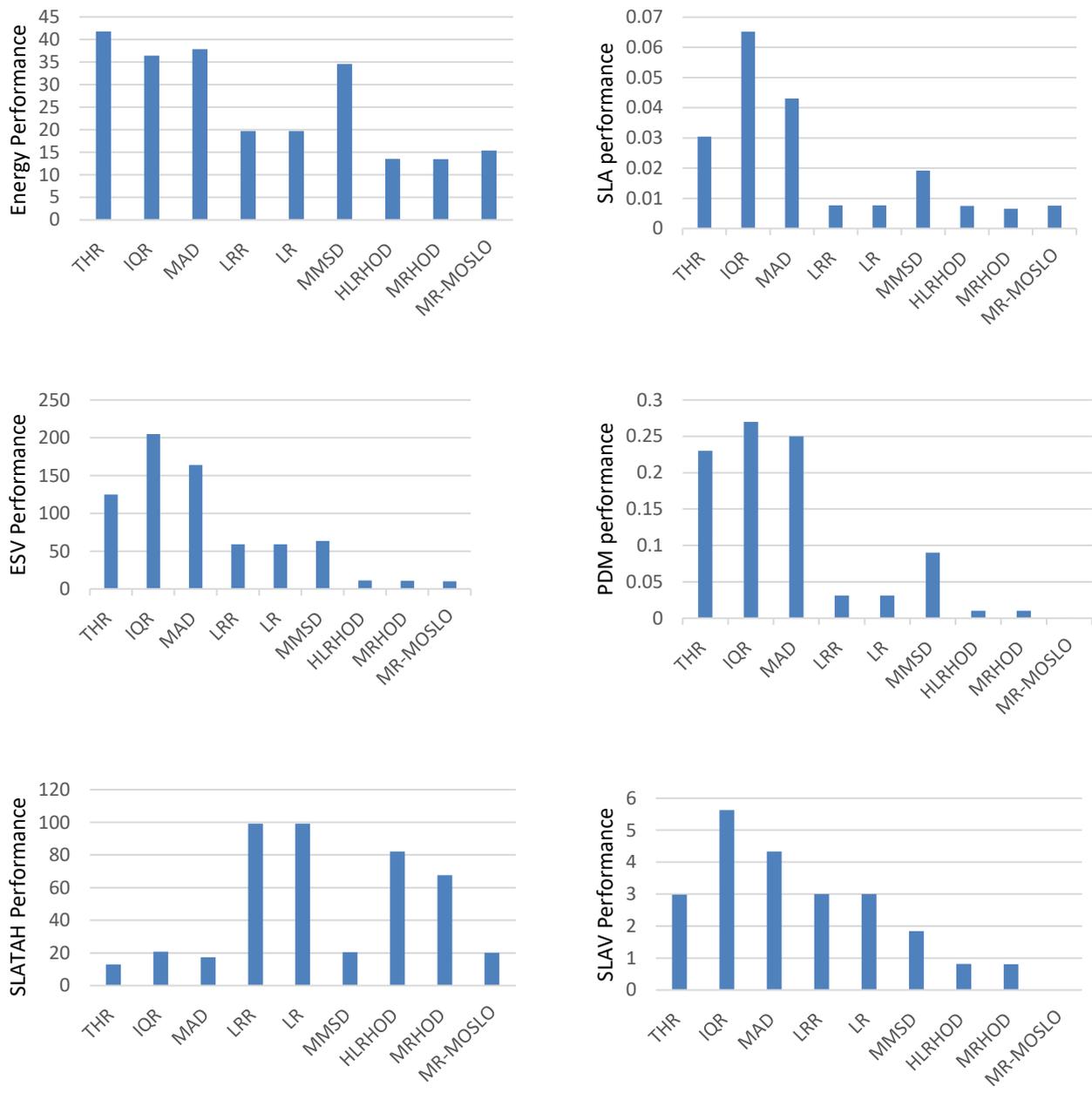

Figure. 1 Algorithms comparative comparison

## 6. Conclusion

Multiple regressions based on Multi-Objective Seven-Spot Ladybird Optimization algorithm for hot overload detection has been developed in this paper. First, the predicted utilization of the host is computed by multiple regressions based on the CPU, RAM and BW utilizations. This increases the effectiveness of host utilization prediction due to the consideration of all three major resources in cloud data centers. Then to determine the host overload status, the adaptive thresholds are estimated optimally using the MOSLO based on host utilizations in number of iterations. The obtained utilizations are ranked in descending order with the three fitness parameters CPU, RAM and BW where CPU is prioritized for sorting the ranking order. The highest utilization (less than 1) and smallest utilization (greater than 0) are selected as upper and lower thresholds. Depending on these thresholds, the host overload status is determined as overload, under-load or normal loaded. Experiments results demonstrated that the proposed MR-MOSLO host overload detection algorithm improves the VM consolidation. For 25 hosts, 30 VMs and 500 tasks, MR-MOSLO achieved lesser values of SLATAH of 20.0434, PDM of 8.7E-4, SLAV of 3.7E-5 and ESV of 10.962 than the other methods. The other parameters namely energy of 15.4 kWh and SLA of 0.00757 are slightly higher than MRHOD and HLRHOD; however they are negligible considered that the overall performance is significant. As a future work, utilizing the new resource parameters other than CPU, RAM and BW will be investigated to further enhance the performance in terms of energy and SLA. Enhancing the performance of other VM consolidation processes of VM selection and placement is also an interesting direction of future work.